\documentclass[twocolumn,floatfix, nofootinbib,prd,preprintnumbers,showpacs,showkeys,superscriptaddress,longbibliography]{revtex4-1}

\usepackage[mathscr]{euscript}
\usepackage{amsmath}
\usepackage{graphicx}
\usepackage{dcolumn}
\usepackage{bm}
\usepackage{epsfig}
\usepackage{amssymb,latexsym,mathrsfs}
\usepackage{graphicx}
\usepackage{color}
\usepackage{hyperref}
\usepackage{float}
\usepackage[thinlines]{easytable}
\usepackage{multirow}
\usepackage{diagbox}

\newcolumntype{M}[1]{>{\centering\arraybackslash}m{#1}}
\newcolumntype{N}{@{}m{0pt}@{}}

\usepackage{amsmath}
\newcommand*\diff{\mathop{}\!\mathrm{d}}

\hypersetup{
    colorlinks=true,
    linkcolor=red,
    citecolor=blue,
}

\usepackage{amsmath}
\usepackage{amssymb}
\usepackage{subfigure}
\usepackage{hyperref}
\usepackage{url}
\usepackage{xcolor}
\usepackage{color}
\definecolor{amaranth}{rgb}{0.9, 0.17, 0.31}
\definecolor{purple(munsell)}{rgb}{0.62, 0.0, 0.77}
\definecolor{americanrose}{rgb}{1.0, 0.01, 0.24}
\definecolor{palatinateblue}{rgb}{0.15, 0.23, 0.89}
\definecolor{royalblue(web)}{rgb}{0.25, 0.41, 0.88}
\definecolor{hanpurple}{rgb}{0.32, 0.09, 0.98}
\definecolor{beaublue}{rgb}{0.74, 0.83, 0.9}
\definecolor{carminered}{rgb}{1.0, 0.0, 0.22}
\definecolor{brightpink}{rgb}{1.0, 0.0, 0.5}
\definecolor{vividviolet}{rgb}{0.62, 0.0, 1.0}

\definecolor{electron}{rgb}{1.0, 0.67, 0.22}
\hypersetup{ linktoc=all,
    colorlinks, linkcolor={palatinateblue},
    citecolor={brightpink}, urlcolor={amaranth}}

\newcommand{\be}{\begin{equation}}
\newcommand{\ee}{\end{equation}}
\newcommand{\bs}{\begin{split}} 
\newcommand{\bea}{\begin{eqnarray}}
\newcommand{\eea}{\end{eqnarray}}

\newcommand{\bes}{\begin{subequations}}
\newcommand{\ees}{\end{subequations}}

\newcommand{\bo}{\raise-1mm\hbox{\Large$\Box$}}

\newcommand{\bd}{\boldsymbol}

\usepackage{physics}
\usepackage{csquotes}
\usepackage{mathtools}
\begin{document}

%\title{On the temperature of lowest order inner bremsstrahlung}
%\title{On the temperature of electron radiation}
%\title{Beta particle trajectory}
%\title{What is the temperature of an accelerating electron?}
\title{Thermal Larmor radiation}
%\title{Classical thermal radiation from an accelerating electron}
%\title{On radiation from beta decay, black holes and flying mirrors}
%\title{Black hole and flying mirror for beta decay}
%\title{Black mirror analog for beta decay}
%\title{Infrared acceleration radiation}
%\title{Acceleration radiation for beta decay}
%\title{Accelerated boundary for beta decay}
%\title{Quantum power for violent acceleration}
%\title{Beta decay acceleration radiation}
%via a black hole remnant analog moving mirror }
%\title{Black hole remnant analog moving mirror model applied to the bremstrahlung radiation during beta decay to find the power and time-dependent angular distribution}
%\title{Radiation from violent acceleration}%
%\title{Acceleration of the electron during beta decay}
%\title{Relativistic quantum radiation from violent acceleration}
%\title{Relativistic quantum acceleration radiation from beta decay}
%\title{Electron mirrors}
%\title{Soft clouds and moving mirrors}
%\title{Undressing in front of a moving mirror}
%\title{Time and temperature of soft clouds, moving mirrors, and inner bremsstrahlung}
%\title{Soft radiation by violent acceleration}
%\title{Acceleration for infrared radiation}
%\title{Infrared emission from an accelerated electron}
%\title{Infrared emission from a rapidly accelerated relativistic electron}
\author{Evgenii Ievlev}
\email{evgenii.ievlev@nu.edu.kz}
\altaffiliation[On leave of absence from ]{National Research Center “Kurchatov Institute”, Petersburg Nuclear Physics
Institute, St.\;Petersburg 188300, Russia}
\affiliation{Physics Department \& Energetic Cosmos Laboratory, Nazarbayev University,\\
Astana 010000, Qazaqstan}
\affiliation{Theoretical and Nuclear Physics Department, al-Farabi Qazaq National University,\\ 
Almaty 050040, Qazaqstan}
\author{Michael R.R. Good}
\email{michael.good@nu.edu.kz}
\affiliation{Physics Department \& Energetic Cosmos Laboratory, Nazarbayev University,\\
Astana 010000, Qazaqstan}
\affiliation{Leung Center for Cosmology and Particle Astrophysics,
National Taiwan University,\\ Taipei 10617, Taiwan}

\begin{abstract} 
%Classical thermal radiation from a single relativistically accelerating electron is found.    
Thermal radiation from a moving point charge is found. The calculation is entirely from a classical point of view, but is shown to have an immediate connection to quantum field theory.
\end{abstract} 

%\keywords{moving mirrors, beta decay, black hole evaporation, acceleration radiation, dressed electrons}
%04.70.Dy (Quantum aspects of black holes)}
%\pacs{04.62.+v, 03.67.Hk, 04.70.-s}
\date{\today} 

\maketitle

%\tableofcontents

%%%%%%%%%%%%%%%%%%%%%%%%%%%%%%%%%%%%%%%%%%%%%
\tableofcontents
%\section{Introduction}

%\section{Classical Canonical Derivation}
\section{Introduction}
\subsection{Motivation}
Thermal Larmor radiation is a fundamental classical phenomenon in which a single accelerating electron emits electromagnetic radiation in accordance with a Planck curve. As such, this process, which to our knowledge has not been derived before, is of paramount importance for understanding a wide range of physical phenomena to which black body emission applies: from the behavior of single moving point charges to the thermal properties of fluctuation–dissipation in the unavoidable white noise of a resistor.

With its far-reaching implications, such as the connection to the quantum understanding of the acceleration-temperature relation, studying thermal Larmor radiation promises to yield new experimental insights into the nature of black hole evaporation \cite{Hawking:1974sw}, moving mirrors \cite{DeWitt:1975ys,Davies:1976hi,Davies:1977yv} and acceleration radiation \cite{Fulling:1972md,Davies:1974th,unruh76}. This simple system is an endeavor with important observational implications that offers an exciting opportunity for the research community to further the collective knowledge of the fundamental origin of the link between acceleration and temperature. 

The blackbody energy spectrum for thermal radiation is significant in the history of physics not because of its own appeal, but because of its major effect in exposing the inadequacy of classical theory and in presenting the formulation of quantum mechanics. However, quantum theory need not always be employed in regimes where classical theory can do the trick. In the case of the Planck radiation law we contend that classical theory alone provides insight into the connection between acceleration and temperature via a direct derivation of moving point charge thermal electromagnetic radiation. Our computation complements Boyer's classical derivation of the black body spectrum \cite{Boyer:1969zz}.  However, presented here, we require no assumption of zero-point radiation, dipole oscillators, or equal partition.  Our classical radiation result is an interesting illustration that even a moving point charge can radiate like a black body without any notion of discrete or discontinuous processes.%; exposing a previous unknown demonstration of the link between acceleration and temperature.

\subsection{Orientation}
We will classically compute the temperature, $T$, of the radiation from an accelerating electron, along with the spectral angular distribution $\diff I(\omega)/d\Omega$, and the spectrum $I(\omega) = \diff E/\diff \omega$. All together the results, in natural units, $\hbar = \mu_0 = c = 1$, with charge $e$, are:
%\be T_{\textrm{electron}} = \frac{\hbar \kappa}{2\pi c k_B},\ee
\begin{equation}
	T_{\textrm{electron}} = \frac{\kappa}{2\pi}
\label{Tnatural}
\end{equation}
%\be \frac{\diff I(\omega)}{\diff \Omega}  = \frac{ \mu _0 c e^2 s^2 \sin ^2\theta }{8 \pi ^2  (c-s \cos \theta)^2}\frac{c \omega/\kappa }{e^{2 \pi  c \omega/\kappa}-1},\label{dIdOmega}\ee
\begin{equation}
	\frac{\diff I(\omega)}{\diff \Omega}  = \frac{ e^2 s^2 \sin ^2\theta }{16 \pi ^3  (1-s \cos \theta)^2}\frac{2\pi \omega/\kappa }{e^{2 \pi   \omega/\kappa}-1},
\label{dIdOmega}
\end{equation} 
%\be I(\omega) = \frac{\mu_0 c e^2}{2\pi^2}\left(\frac{c \eta}{s}-1\right) \frac{2\pi c \omega/\kappa}{e^{2\pi c \omega/\kappa}-1}.\label{Iw}\ee
\be I(\omega) = \frac{e^2}{2\pi^2}\left(\frac{\eta}{s}-1\right) \frac{2\pi \omega/\kappa}{e^{2\pi  \omega/\kappa}-1}.\label{Iw}\ee
Here $\kappa>0$ is the acceleration parameter defining the trajectory of the electron; it is not the proper acceleration (more on this later).  The final speed of the electron is $0<s<1$, $\omega$ is the frequency of the light, and $\eta = \tanh^{-1} s$ is the final rapidity of the electron.  The polar angle is $0\leq\theta \leq \pi$; while the electron travels rectilinearly in the radial $\hat{\bd{r}}$ direction. It should be noted, in these units, that $e^2 = 4\pi \alpha$ where $\alpha$ is the fine structure constant.

%\subsection{Consistency Checks}
%Consistency is shown by agreement with the energy,
%\be E = \frac{\kappa \mu_0 e^2}{24\pi}\left(\frac{c\eta}{s}-1\right),\ee
%the deep infrared spectrum,
%\be I_{\textrm{infra}} = \frac{\mu_0 c e^2}{2\pi^2}\left(\frac{c\eta}{s}-1\right),\ee
%and the moving mirror beta Bogoliubov spectrum,
%\be \frac{|j_\alpha(k)|^2}{(2\pi)^2} = |\beta_{pq}|^2,\ee 
%as well as the associated Planck factor thermality.

\subsection{Discussion}

There have been several studies investigating the classical connections between acceleration and temperature; most of which deal with uniform proper accelerated trajectories. For instance, Cozzella et al \cite{Cozzella:2017ckb} claim the observation of classical Larmor radiation is a signal of the quantum Davies-Fulling-Unruh effect, suggesting that a quantum effect can be verified through a classical computation.  They use a uniformly accelerated charge and argue the Unruh thermal bath is codified in the Larmor radiation \cite{Cozzella:2020gci} emitted from the accelerated charge.

Leonhardt et al \cite{Leonhardt:2017lwm} developed a water-wave model with the boundary of the container acting like a mirror, revealing a classical notion of the Unruh effect as the correlation of noise in space and time.  There they replaced $\hbar$ by the strength of classical noise and $c$ by the speed of the waves involved in the effect. They suggest the use of non-uniform accelerated trajectories could help extend the idea to the quantum regime.  

Recently, Hegelich et al \cite{Hegelich:2022zca} relate the model detector of the Davies-Fulling-Unruh temperature to experiment by considering a point-like electron in its place.  Since the electron has no internal degree of freedom, thermalization of this uniformly accelerated `detector' occurs in a unexpected way: fluctuations in the plane transverse to the acceleration are amplified into radiated particles.  

In fact, when considering accelerated detectors with sufficient thermalization time in the limit of no internal structure \cite{Cozzella:2020gci}, the Davies-Fulling-Unruh power reduces to the classical Larmor power formula for accelerated electrons, \cite{Lynch:2019hmk}; see also the Multi-Petawatt Physics Prioritization (MP3) Workshop Report \cite{DiPiazza:2022wzo}. This is a particularly suggestive result connecting Larmor radiation with the Davies-Fulling-Unruh effect. 

In an expanded version of the Gregori et al \cite{Gregori:2023tun} presentation at the MP3 workshop \cite{DiPiazza:2022wzo}, it is emphasized that the Davies-Fulling-Unruh effect is a very general process, associated with all accelerated bodies, regardless of the presence of an event horizon; supporting the importance of studying horizonless and asymptotic inertial trajectories.  In particular, Gregori et al. stress that to confirm the Unruh effect, a framework is needed where the lab acceleration is not constant.  This is the approach we take with an electron which has a non-uniform proper acceleration. 

 Morever, Gregori et al \cite{Gregori:2023tun} use a heuristic derivation of the Unruh effect whose realization is dependent on quantum discreteness.  They highlight the use of the electron as a detector with no internal structure; calling careful attention to how to define and register a change of state. Specifically they note that this issue is seldom discussed and how controversial \cite{Ford:2005sh} the issues are in distinguishing the Unruh effect from other classical and quantum radiation processes when involving accelerating moving point charges.  Our classical approach confronts these important points without the need for discrete energy levels. 

 Others argue similarily; for example, Pauri and Vallisneri \cite{Pauri:1999nu} contend that the Unruh effect is deeply rooted at the classical level and could have been predicted earlier and by a different route, drawing from the analogy of radiation in classical electromagnetism.  Lin \cite{Lin:2001hw} found the vacuum expectation value of the energy density for a point-like electron is identical to its classical self-energy density, pushing for a classical correspondence of the Unruh effect. 

Beyond the classical connections, such as those found by Boyer \cite{Boyer:1984yqq} and Cole \cite{Cole:1987pw}, who discussed the thermal effects of acceleration within classical theory including classical zero-point radiation, there is good motivation for studying non-uniformly accelerated motions.  Boyer found it seemingly discouraging that in the classical electromagnetic case, an observer undergoing uniform acceleration \cite{Boyer:1980wu}, through classical electromagnetic zero-point radiation, detects field correlation functions corresponding to a spectrum different from Planck's spectrum.

In much the same way, the present authors are not alone in being perplexed by the non-Planckian spectrum resulting from straightforward assumption of proper uniform acceleration, e.g. \cite{good2020extreme}.  In our view, the fact that a constant proper accelerated moving charge does not emit thermal radiation underscores the importance of studying alternative trajectories and potential connections to the Planck spectrum. 

The aforementioned moving mirror model is well-suited to non-uniform motions and wide variety of applications.  For instance, the moving mirror has recently been applied to address finite-size \cite{Lin:2021bpe,Lin:2022sbi} and model entanglement \cite{Reyes:2021npy,Akal:2020twv}. By applying this simple conceptual analog to the study of radiation emitted by an electron, we develop a connection between the electron and mirror.

 An early clue this correspondence existed was perhaps first recognized via radiation reaction by Ford-Vilenkin \cite{Ford:1982ct} and substantially established by Nikishov-Ritus \cite{Nikishov:1995qs}. Further development on Bogolyubov-current connection occurs in Ritus \cite{Ritus:1999eu,Ritus:2002rq,Ritus:2003wu,Ritus:2022bph}.  The correspondence to Larmor power was derived in Zhakenuly et al \cite{Zhakenuly:2021pfm}. The connection and recipe we develop here is general enough to be applied to any integrable classical trajectory that emits finite radiation energy but also precise enough to directly calculate the relevant integrable spectra for specified electron accelerations. We explicitly do this and show that the approach is consistent. In this analog, the electron is treated as a moving mirror, similar to the way black holes are treated as moving mirrors, e.g. Schwarzschild \cite{Good:2016oey}, Reissner–Nordstr\"om \cite{good2020particle}, and Kerr \cite{Good:2020fjz}, but with limited acceleration; i.e. asymptotic inertia characterized by a proper acceleration $\mathcal{A} \to 0$. The specific solution we focus on behaves as a remnant presciently described in Wilczek \cite{wilczek1993quantum}. The generality in this paper reveals a previously unknown thermal electron acceleration radiation spectra, which can help to establish links between acceleration, gravity, and thermodynamics in a precisely applied manner.

There is good reason to study the one-channel Planck curve, Eq.~(\ref{Iw}).  It is important because its fundamental thermal radiation is closely connected to Johnson-Nyquist (white) noise \cite{nyquist}.  A hot resistor produces electromagnetic waves along its single mode, transmitted in one spatial dimension along the line just as a hot body will produce electromagnetic waves isotropically in free space, see e.g. \cite{oliver}. One dimensional white noise power from a resistor is equal to the power picked up by an antenna pointed at a three dimensional blackbody at the same temperature \cite{Dicke}. Beyond basic thermodynamics, white noise has applications in applied electronics, metrology, material science, signal processing, and telecommunications. For instance, in microwave photonics, the one-channel Planck curve provides a theoretical framework for understanding and mitigating the impact of thermal noise on electronic systems \cite{urick2021fundamentals}; in particular at high frequencies or low temperatures due to quantum effects.

\section{Larmor Acceleration and Temperature}
\subsection{Prelude to Larmor Equilibrium}

Larmor \cite{Larmor1897} found the non-relativistic formula for the total power emitted by an accelerating point charge. Written in SI units \cite{Griffiths:1492149},
\be P = \frac{2}{3} \frac{e^2}{4\pi \epsilon_0 c^3} a^2 = \frac{\mu_0 e^2}{6\pi c} a^2,\ee
where the vacuum magnetic permeability, $\mu_0$, and the vacuum electric permittivity, $\epsilon_0$, are related to the speed of light by $\mu_0 \epsilon_0 = 1/c^{2}$.  Here $e$ is the electric charge and $a = \dot{v}$ is the non-relativistic acceleration.

At low speeds, and for limited motions with asymptotic inertia, the total energy emitted is 
\be E = \int_{-\infty}^{\infty} P \diff{t}.\ee
In equilibrium, a uniformly accelerated point charge might be expected to radiate thermally with constant power.  However, one can see eternal constant acceleration will not give a global finite energy.   

Recently, it has been shown that a specific relativistic trajectory \cite{Good:2016yht} with inverse velocity, $v$, final speed $s$, and free parameter $\kappa$ (with units of acceleration),
\begin{equation}
	\frac{1}{v} = \frac{c}{\kappa z} + \frac{1}{s},
\label{velocity}
\end{equation}
emits a total finite energy \cite{Good:2022eub},
\be E = \frac{\mu_0 e^2 \kappa}{24\pi}\left(\frac{\tanh^{-1}s/c}{s/c}-1\right),\label{TE0}\ee
consistent with a finite period of constant power \cite{Good:2022xin},
\be \bar{P}_c = \frac{\mu_0 e^2 \kappa^2}{48\pi c}.\ee
In turn, this equilibrium emission is associated with uniform local\footnote{This is also called the `peeling function', see e.g. \cite{Bianchi:2014qua,Barcelo:2010pj}. We will call it the peel acceleration or `peel' for short.} acceleration \cite{CW2lifetime},
\be \bar{\kappa}(u) = \frac{v''(u)}{v'(u)} \rightarrow \kappa,\ee
where $v(u)$ is the advanced trajectory, $v = t+r$, in terms of retarded time, $u = t-r$, using light-cone coordinates.  In the following section, we demonstrate this trajectory emits a thermal radiation spectrum throughout its motion. 

%Let us then consider the covariantly written, relativistic Larmor formula \cite{Jackson:490457},
%\be P = \frac{\mu_0 e^2}{6\pi c}\alpha^2,\ee
%where the non-relativistic acceleration $a = \dot{v}$ has been replaced by the proper acceleration, which for rectilinear motions is $\alpha = \gamma^3 a$.% In addition, let us integrate over retarded time $u= t-r$, so that,
%\be E = \int_{-\infty}^{\infty} \frac{\diff E}{\diff u} \diff u = \int_{-\infty}^{\infty}  \frac{\mu_0 e^2}{6\pi c}\alpha^2 \frac{\diff t}{\diff u} \diff u,\ee
%where
%\be \bar{P} =  \frac{\diff E}{\diff u} = \frac{\mu_0 e^2}{6\pi c}\alpha^2 \frac{1}{1-v}.\ee
\subsection{Classical Thermal Spectrum Derivation}
A classical spectrum is obtained for the radiation emitted by the moving point charge, demonstrating a Planck distribution.
This computation will result in the spectrum, $I(\omega) = \diff E/\diff \omega$ by first computing the angular distribution, $\diff I(\omega)/\diff \Omega \equiv \diff^2E/\diff \omega \diff\Omega$. 

\begin{figure}[htbp]
\centering
%\begin{subfigure}{0.5\textwidth}
  \centering
  \includegraphics[width=0.8\linewidth]{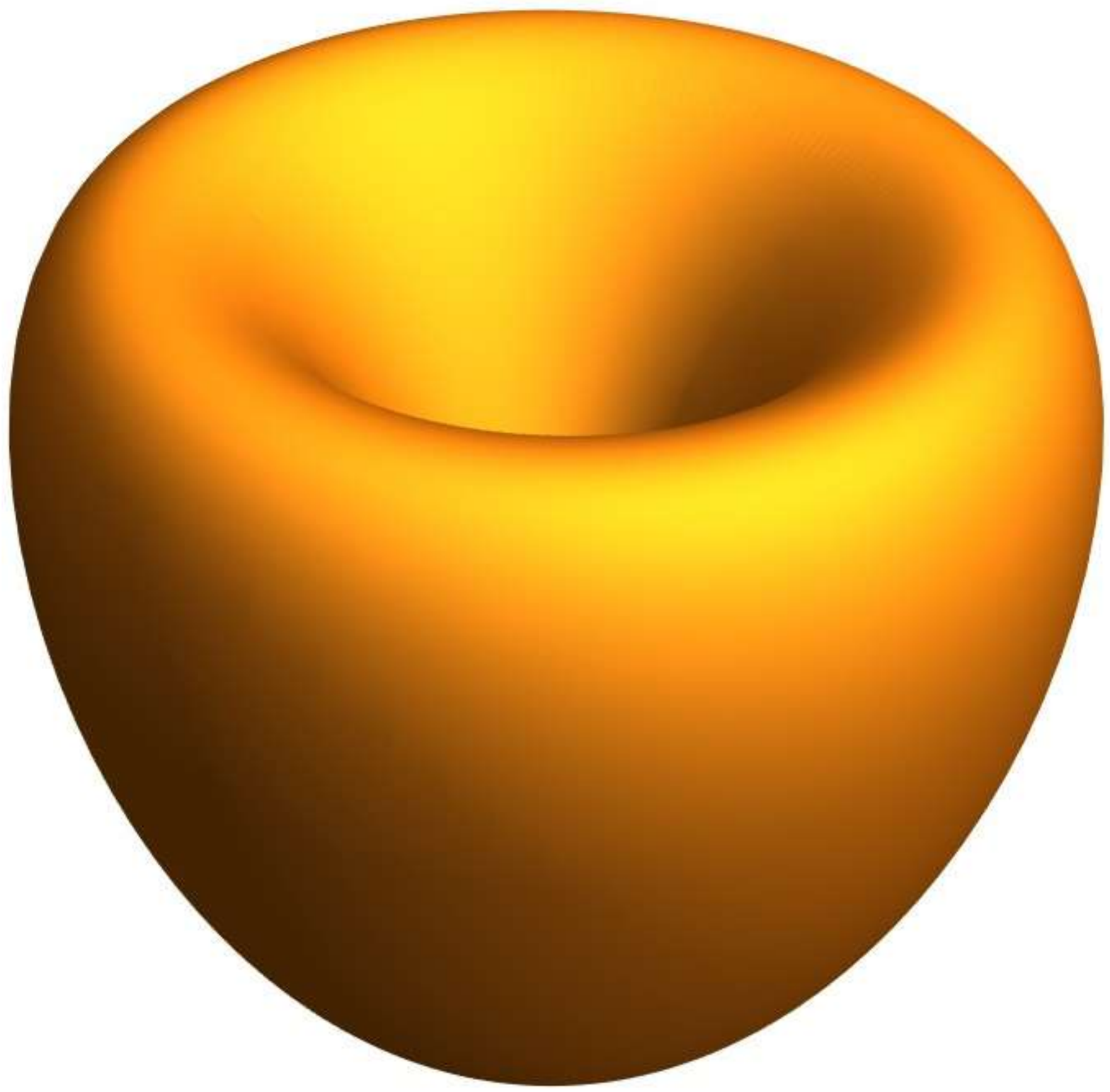}
 \caption{A plot of Eq.~(\ref{dI_dOmega}), the distribution, $\diff I/\diff \Omega$.  Here we use unit charge, natural units and $\omega = \kappa = 1$. The final speed of the electron is $s= 0.9$. The spectral distribution is that of rectilinear allocation peaking about the forward direction but not actually in the forward direction. }  
\label{Fig_distribution}
\end{figure}

\begin{figure}[htbp]
\centering
%\begin{subfigure}{0.5\textwidth}
  \centering
  \includegraphics[width=1.0\linewidth]{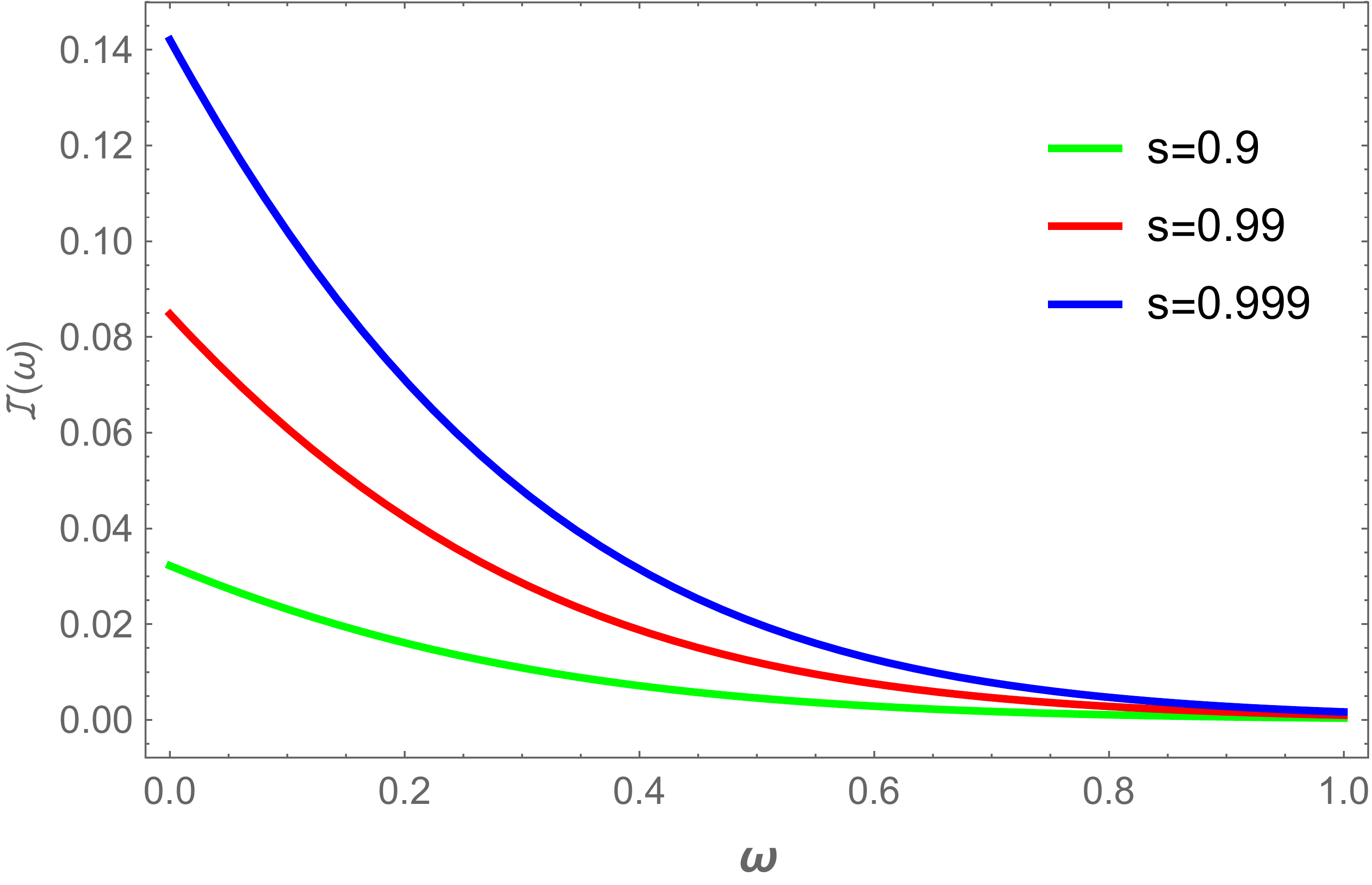}
 \caption{A plot of Eq.~(\ref{I_result_1}), the spectrum, $I(\omega)$.  Here we use unit charge, natural units and $\kappa = 1$. The final speed of the electron is $s= 0.9$, $s=0.99$, and $s=0.999$ for green, red, and blue, respectively. This plot illustrates the shape of 1-D black body radiation with peak deep infrared color, $\omega =0$}  
\label{Fig_blackbody}
\end{figure}

For clarity, SI units are employed and we start with the general radiation spectrum of a moving point charge, 
(see, SI units e.g. Eq. 23.89 p. (911) of Zangwill \cite{Zangwill:1507229}
or Gaussian units Eq. (14.67) p. (701) of Jackson \cite{Jackson:490457}): 
\begin{equation}
	\frac{\diff I(\omega)}{\diff \Omega} = \frac{\mu_0 e^2\omega^2}{16\pi^3c}\left|\bd{\hat{n}} \times \int\displaylimits_{-\infty}^{\infty} \diff t\, \bd{\beta}(t) e^{i\phi}\right|^2.
\label{density_zangwill}
\end{equation}
where $\phi =\omega t -\bd{k}\cdot\bd{r}(t)$, while $\bd{\hat{n}}$ is the direction of $\bd{k}$. The trajectory function is 
\be \bd{r}(t) = \frac{sc}{\kappa}W(e^{\kappa t/c})\bd{\hat{r}},\ee
where $W$ is the Lambert product logarithm, and $\bd{\hat{r}}$ is a unit vector in the movement direction (we choose it to be along the $z$-axis).
The dimensionful quantity $0<s<c$ is the final speed of the electron. Using $\bd{k} = (\omega/c) \bd{\hat{n}}$ and simplifying the cross product,
\be\frac{\diff I(\omega)}{\diff \Omega} = \frac{\mu_0 e^2\omega^2}{16\pi^3c}\left|\sin\theta \int\displaylimits_{-\infty}^{\infty} \diff t\, \dot{z}(t) e^{i\phi}\right|^2,\ee
where $\phi = \omega (t - [z(t)/c]\cos\theta)$. Inverting the trajectory to remove the productlog gives
\begin{equation}
	t(z) = \frac{c}{\kappa}\ln \left(\frac{\kappa z}{s c}\right)+\frac{z}{s},
\label{traj_t}
\end{equation}
and integrating over $z$ rather than $t$ we obtain
\begin{equation}
	\frac{\diff I(\omega)}{\diff \Omega} = \frac{\mu_0 e^2\omega^2}{16\pi^3c}\left|\sin\theta \int\displaylimits_{0}^{\infty} \diff z  e^{i \phi(z)}  \right|^2,
\label{I_phi_int}
\end{equation}
with
\begin{equation}
	e^{i \phi(z)} = \left(\frac{\kappa  z}{c s}\right)^{\frac{i c \omega }{\kappa }} e^{i \frac{z \omega}{s} \left( 1 - \frac{s}{c} \cos\theta \right) }.
\end{equation}
%%
%To compute the integral $\int\displaylimits_{0}^{\infty} \diff r  e^{i \phi(r)}$ we pass to the variable $\rho$ defined%
%\footnote{\label{ftnt:log}  The integrand as a function of complex variable is, generally speaking, a multivalued function. To resolve any ambiguities we write here explicitly $e^{i \pi/2}$ instead of just $i$. This can be traced back the logarithm in Eq.~\eqref{traj_t}, for which we choose the usual Riemann sheet with the cut going from $0$ to $-\infty$. } 
%as
%\begin{equation}
%	r = \rho \cdot e^{i \frac{\pi}{2}} \frac{s}{ \omega \left( 1 - \frac{s}{c} \cos\theta \right) } \,.
%\end{equation}
%%
%and rotate the $\rho$-contour to the positive real axis. This yields the integral of the $\Gamma$-function form, and the result is
%\begin{equation}
%	\Gamma \left(\frac{c i \omega }{\kappa }+1\right) \left(\frac{\kappa }{c s}\right)^{\frac{i c \omega }{\kappa }}
%		 \frac{e^{i \frac{\pi}{2} - \frac{\pi c \omega}{2 \kappa}}}{ \left( \frac{\omega}{s} - \frac{\omega}{c} \cos\theta \right)^{1+\frac{i c \omega }{\kappa }} }  \,.
%\end{equation}
%
%Substituting this into Eq.~\eqref{I_phi_int} gives the distribution
%
The integral here is computed in Appendix~\ref{sec:int_calc}, the result is
\begin{equation}
	\frac{\diff I(\omega)}{\diff \Omega}  = \frac{ \mu _0 c e^2 s^2 \sin ^2\theta }{16 \pi^3 (c-s \cos \theta)^2}\frac{2\pi c \omega/\kappa }{e^{2 \pi  c \omega/\kappa}-1}.
\label{dI_dOmega}
\end{equation}
%
%Complex conjugating simplifies to
%\be |...|^2 = \frac{\pi  c^3 s^2 \sin ^2\theta \left(\coth \left(\frac{\pi  c \omega }{\kappa }\right)-1\right)}{\kappa  \omega  (c-s \cos \theta)^2}.\ee
%The angular spectrum is therefore,
%\be \frac{\diff I(\omega)}{\diff \Omega}  = \frac{ \mu _0 c e^2 s^2 \sin ^2\theta }{8 \pi ^2  (c-s \cos \theta)^2}\frac{c \omega/\kappa }{e^{2 \pi  c \omega/\kappa}-1}.\ee
This is the SI version of Eq.~(\ref{dIdOmega}). See Figure \ref{Fig_distribution} for a spherical 3D plot of the distribution Eq.~(\ref{dI_dOmega}). Integration over $\diff \Omega = \sin\theta \diff \theta \diff \phi$ gives 
\begin{equation}
	I(\omega) = \frac{\mu_0 c e^2}{2\pi^2}\left(\frac{c \eta}{s}-1\right) \frac{2\pi c \omega/\kappa}{e^{2\pi c \omega/\kappa}-1}.
\label{I_result_1}
\end{equation}
This is exactly Eq.~\eqref{Iw} in SI units.
Here $\eta = \tanh^{-1} s/c$. This is the SI version of Eq.~(\ref{Iw}). See Figure \ref{Fig_blackbody} for a plot of the black body curve of $I(\omega)$, Eq.~(\ref{I_result_1}). 

%Notice the Planck distribution black body curve has been found without appeal to quantum assumptions. 
%The spectral density Eq.~\eqref{I_result_1} is very similar to the Planck distribution for a black body in 1+1 dimensions. 
%Below we will see that it indeed corresponds to a radiation at a certain temperature. 
The spectral density Eq.~\eqref{I_result_1} is very similar to the Planck distribution for a black body in 1+1 dimensions. 
Indeed, Eq.~\eqref{I_result_1} corresponds to a radiation at a certain temperature. 
We present the formula for the temperature in Eq.~\eqref{Tnatural}, see Sec.~\ref{sec:tale} for more details.

We emphasize that our derivation here was purely classical, and that the spectral density Eq.~\eqref{I_result_1} suggests that the radiation is legitimately thermal \cite{Boyer:1969zz,oliver}.
This means that we see a genuine temperature effect as a result of an acceleration, i.e. the Davies-Fulling-Unruh effect, without any appeal to quantum.

\subsection{A Tale of Two Thermometers}
\label{sec:tale}

We are left with the task of defining a temperature $T$ to conform with the one spatial dimensional blackbody Planck distribution of Eq.~(\ref{I_result_1}),
\be I(\omega) \sim  \frac{2\pi c \omega/\kappa}{e^{2\pi c \omega/\kappa}-1}.\ee
Do we use the pre-existing \cite{barrowSTONE}, classical scale of the problem: $\mu_0 c e^2$? 
\begin{equation}
	\frac{2\pi c \omega}{\kappa} \overset{?}{\leftrightarrow} \frac{\mu_0 c e^2 \omega}{k_B T}. \label{StoneyScale}
\end{equation}
Or do we introduce an additional scale, by route of $\hbar$, which is typically associated with quantum theory?

\be  \frac{2\pi c \omega}{\kappa} \overset{?}{\leftrightarrow} \frac{\hbar \omega}{k_B T}.\label{KelvinScale}\ee
The former could be called the Stoney scale\footnote{If the Stoney scale, Eq.~(\ref{StoneyScale}), were used, $\alpha$ being the fine structure constant, and $\hbar = \mu_0 c e^2 /4\pi \alpha$, the temperature is written
\be \nonumber T = \frac{\mu_0 e^2 \kappa }{2\pi k_B} = 2\alpha \frac{\hbar \kappa}{c k_B} \quad \to \quad T = 2\alpha \kappa.  \ee
The arrow indicating natural units set $\hbar = c = k_B = 1$. } \cite{stoney1881physical} while the latter is the Planck scale with unit of temperature Kelvin. Using the latter, quantum-traditional definition of temperature associated with Planck's constant, Eq.~(\ref{KelvinScale}), we find that the classical Larmor radiation is the quantum Davies-Fulling-Unruh effect, with Planck-distributed emission at temperature,
\be T = \frac{\hbar \kappa}{2\pi c k_B}.\ee
Notice $\kappa$ is the acceleration parameter of the local-uniformly accelerated electron, whose trajectory is
Eq.~\eqref{velocity}.
% $1/v = c/(\kappa r) + 1/s$.  
The dynamical meaning of the constant $\kappa$ is the late retarded time $u\to \infty$, high-speed $s\to c$ double limit of the peel \cite{carlitz1987reflections}: $\kappa(u) = v''(u)/v'(u) = 2\mathcal{A} e^\eta = 2\diff \eta(u)/\diff u$, that is
\be \lim_{u\to\infty} \lim_{s\to c} \bar{\kappa}(u) = \kappa.\ee
Here $\mathcal{A}(u)$ is the proper acceleration, $\eta(u)$ is rapidity, and $\bar{\kappa}(u)$ is the peel acceleration \cite{Bianchi:2014qua,Barcelo:2010pj}.

%To emphasize the classical electrodynamic nature of the temperature, we can substitute $\hbar$ in favor of the fine structure constant, $\alpha$, and electric charge $e$, by use 

%\be \hbar = \frac{\mu_0 c e^2}{4\pi \alpha},\ee

%which gives

%\be T = \frac{\mu_0 e^2 \kappa}{8\pi^2 \alpha k_B}.\ee

\section{Electron-mirror connection}

In this section we are going to relate two systems that are seemingly very different in their nature.
One is a point charge moving along some rectilinear trajectory in 3+1 Minkowski spacetime.
The other is a mirror that moves in 1+1 dimensions.
It turns out that the radiation spectra in these systems are the same, which signifies a deep underlying connection between the two.

\subsection{Setup of the Moving Point Charge}

We consider a point charge with four-current%
\footnote{From this point on we use natural units and set the speed of light $c$, the Planck's constant $\hbar$, 
%the electron charge $e$ 
the vacuum magnetic permeability $\mu_0$
and the Boltzmann's constant $k_B$ to 1.  
In these units the vacuum permittivity $\varepsilon_0 = 1$, while electron charge is a dimensionless number $e^2=4 \pi \alpha \approx 0.092$.
Our choice for the spacetime metric signature is $(+,-,-,-)$, the same as in Jackson \cite{Jackson:490457}. % see ds^2 on p. 528 (553) in Jackson above eq. (11.26)
}
(see, e.g. Jackson Sec. 11.9 \cite{Jackson:490457})
\be j_\mu = (\rho, \bd{j}) \ee
%charge density,
%\be \rho =  \delta^3(\bd{r}-\bd{r}(t)),\ee
%and current density,
%\be \bd{j} =  \bd{v}(t)\delta^3(\bd{r}-\bd{r}(t)).\ee
%%
with the charge and current densities
\begin{equation}
	\rho =  e\, \delta^3(\bd{r}-\bd{r}(t)) \,, \quad
	\bd{j} =  e\, \bd{v}(t)\delta^3(\bd{r}-\bd{r}(t)) \,.
\end{equation}
Moreover, we are going to focus on the case of a \textit{rectilinear} motion along the $z$ axis, so that 
\begin{equation}
	j_x = j_y = 0 \,,
\label{j_xy}
\end{equation}
In this case we have the following formula for the radiated energy spectral density 
(cf. Eq.~\eqref{density_zangwill}, see also Eq. (14.70) of \cite{Jackson:490457}): % Jackson p. 676 (701)
\begin{equation}
	\frac{\diff I(\omega)}{\diff \Omega} = 
%		\frac{\omega^2}{16 \pi^3} \sin^2\theta \, \abs{   j_z(\omega, \bd{\hat{n}} \omega ) }^2 \,.
		\frac{\omega^2}{16 \pi^3} \sin^2\theta \, \abs{   j_z(\omega, k_z ) }^2 \,,
\label{I_Omega}
\end{equation}
%
%Here $\bd{\hat{n}}$ is the unit vector in the direction of observation, and $j_z(\omega, \bd{\hat{n}} \omega )$ is the Fourier transform of the current, taken for the wavevector $\bd{k} = \bd{\hat{n}} \omega$.
where $k_z = \omega \cos\theta$ (for such a rectilinear motion the energy distribution does not depend on the $k_x, k_y$ components of the wavevector).
The total radiated energy is evaluated as an integral of the density:
\begin{equation}
\begin{aligned}
	E_\text{tot} 
		&= \int\limits_0^\infty d\omega \int d\Omega \frac{\diff I(\omega)}{\diff \Omega} \\
		&= \frac{1}{16 \pi^3} \int\limits_0^\infty d\omega  \int d \Omega \ \omega^2 \sin^2\theta \, \abs{   j_z(\omega, k_z ) }^2 \,.
\end{aligned}
\label{Etot_pointcharge}
\end{equation}

\subsection{Connection to the Moving Mirror}
\label{sec:to_the_mirror}

Now consider a 1+1-dimensional moving mirror \cite{DeWitt:1975ys,Davies:1976hi,Davies:1977yv}. We take the mirror moving along the same trajectory as the point charge from the previous subsection.
This is possible since the point charge is in rectilinear motion.

The 1+1 setup can be roughly described as follows, e.g. \cite{Birrell:1982ix}.
Consider a massless scalar with the standard Lagrangian.
In a trivial Minkowski vacuum the spacetime is homogeneous, and there is no radiation.
However when one introduces a moving mirror (dynamical Casimir effect), the vacuum is perturbed as there are non-trivial boundary conditions now \cite{Good:2021asq}.
The mirror causes creation of scalar field quanta which are seen as radiation.
Total radiated energy for a mirror is given by the formula, e.g. \cite{walker1985particle},
\begin{equation}
	E_\text{tot}^\beta = \int\limits_0^\infty dp \int\limits_0^\infty dq \ p\; |\beta_{pq}|^2  \,,
\label{Etot_betas}
\end{equation}
where $\beta_{pq}$ is the beta Bogolubov coefficient \cite{Parker:2009uva}; with frequency modes $q$ for incoming plane wave form and $p$ for outgoing plane wave form, see e.g. \cite{Fabbri}.  Here the beta coefficient accounts is the sum of the squares for each side of the perfectly reflecting mirror \cite{Zhakenuly:2021pfm,Good:2022gvk,Good:2023hsv}.

Following Nikishov-Ritus \cite{Nikishov:1995qs}, we want to relate $|\beta_{pq}|^2$ to some quantity characteristic to the moving point charge, see also \cite{Ritus:1999eu,Ritus:2002rq,Ritus:2003wu,Ritus:2022bph}.
Since $|\beta_{pq}|^2$ corresponds to a one-dimensional system, we can expect that the corresponding quantity in the 3+1 system should be a Lorentz scalar.
From the four-current $j_\mu$ we build a scalar $- j_\mu j^\mu = \bd{j}^2 - \rho^2 $. 
In the Fourier space we consider the corresponding quantity
\begin{equation}
	|j_\mu (\omega, k_z)|^2 \equiv |\bd{j}(\omega, k_z)|^2 - |\rho(\omega, k_z)|^2 \,.
\label{j_scalar}
\end{equation}
In Eq.~(\ref{j_scalar}), $\omega$ is understood as a Fourier transform parameter and can take positive as well as negative values. 
It is customary to take the physical frequency $\omega$ as a positive quantity since the sign of the frequency has no physical meaning. 
This is also motivated by the fact that in calculating the total energy we have to integrate over all frequencies, and we can just as well take the even part of the distribution and integrate over non-negative frequencies only, cf. Sec.~14.5 of \cite{Jackson:490457}.

To follow this convention instead of $|j_\mu (\omega, k_z)|^2$  we should consider the sum 
\begin{equation}
	|j_\mu (\omega, k_z)|^2 + |j_\mu (-\omega, k_z)|^2 \,.
\label{j_scalar_omegapos}
\end{equation}
In this section we are going to check that Eq.~\eqref{j_scalar_omegapos} and $|\beta_{pq}|^2 $ are indeed related to each other.
In fact, they turn out to be proportional to each other.
We present the following formula for this correspondence:
\begin{equation}
	\frac{|j_\mu (\omega, k_z)|^2 + |j_\mu (- \omega, k_z)|^2}{(2\pi)^2 e^2 } = |\beta_{pq}|^2 \,.
\label{beta_j_general}
\end{equation}
Note that the proportionality coefficient depends on the conventions for the Fourier transform.
For our convention see Eq.~\eqref{fourier_def} below.

% ??? Eq.~(7) of \cite{Ritus:2003wu}

To support this proposed formula we are going to check that the formulas for the radiated energy in the point particle case and for the moving mirror give the same result.
We will start from the total energy, see Eq.~\eqref{Etot_pointcharge} and Eq.~\eqref{Etot_betas}. 

As it turns out the integrals in Eq.~\eqref{Etot_pointcharge} and Eq.~\eqref{Etot_betas} are transformed into one another by a change of integration variables.
We show this by transforming the integral in Eq.~\eqref{Etot_betas}.
First let us pass to the light-cone coordinates in frequency space:
\begin{equation}
	2p = \omega + k_z \,, \quad 
	2q =\omega - k_z ,
\label{pq_def}
\end{equation}
%
%Let us take the integral Eq.~\eqref{Etot_betas} and pass from the integration variables $p, q$ to $\omega, k_z$.
The Jacobian is $1/2$, and we have (note the integration limits)
\begin{equation}
\begin{aligned}
	e^2 E_\text{tot}^\beta 
		&= \int\limits_0^\infty \diff \omega \int\limits_{- \omega}^{+ \omega} \diff k_z \frac{1}{2} \frac{\omega + k_z}{2}
				\frac{|j_\mu (\omega, k_z)|^2 + |j_\mu (-\omega, k_z)|^2}{(2\pi)^2} \\
%		&= \frac{1}{16\pi^2} \int\limits_0^\infty \diff \omega \int\limits_{- \omega}^{+ \omega} \diff k_z  ( \omega + k_z ) 
%				(|j_\mu (\omega, k_z)|^2 + |j_\mu (\omega, -k_z)|^2 ) \\
		&= \frac{1}{16\pi^2} \int\limits_0^\infty \diff \omega \int\limits_{- \omega}^{+ \omega} \diff k_z  \omega 
				(|j_\mu (\omega, k_z)|^2 + |j_\mu (\omega, -k_z)|^2 ) \\
		&= \frac{1}{8\pi^2} \int\limits_0^\infty \diff \omega \int\limits_{- \omega}^{+ \omega} \diff k_z \ \omega  \
				|j_\mu (\omega, k_z)|^2 \,.		
\end{aligned}
\label{Etot_beta_transf}
\end{equation}
In the second line we used $j_\mu (-\omega, k_z) = j_\mu (\omega, -k_z)^*$, which follows from the fact that the four-current in the coordinate space is real-valued.
Next we can compare to the point charge formula Eq.~\eqref{Etot_pointcharge}. 
In the latter equation only waves with $|\bd{k}|=\omega$ contribute to the integral.
This helps us to figure out the next change of variables: in Eq.~\eqref{Etot_beta_transf} we pass from $\omega, k_z$ to $\omega, \theta$ according to 
$k_z = |\bd{k}| \cos\theta = \omega \cos\theta$.
The Jacobian is $\omega \sin\theta$, and we have
\begin{equation}
\begin{aligned}
	e^2 E_\text{tot}^\beta 
		&= \frac{1}{8\pi^2} \int\limits_0^\infty \diff \omega \int\limits_{-1}^{+1} \diff (\cos\theta) \ \omega^2  \
				|j_\mu (\omega, \omega\cos\theta)|^2 \\		
%		&= \frac{1}{8\pi^2} \int\limits_0^\infty \diff \omega \int\limits_{-1}^{+1} \diff (\cos\theta) \ \omega^2  \
%				(1 - \cos^2\theta) |j_z (\omega, \omega\cos\theta)|^2 \\
		&= \frac{1}{16\pi^3} \int\limits_0^\infty \diff \omega \int \diff \Omega \ \omega^2  \
				\sin^2\theta |j_z (\omega, \omega\cos\theta)|^2 \,.		
\end{aligned}
\label{Etot_beta_transf_2}
\end{equation}
Here we used the charge conservation law
\begin{equation}
%	\partial_t \rho(t,\bd{r}) + \div \bd{j}(t,\bd{r}) = 0
%	\leftrightarrow
	\omega \rho (\omega, \bd{k}) - \bd{k} \cdot \bd{j} (\omega, \bd{k}) = 0 \,,
\label{cont_eq}
\end{equation}
which together with Eq.~\eqref{j_xy} allows us to express our Lorentz scalar from Eq.~\eqref{j_scalar} as
\begin{equation}
	|j_\mu|^2 
		= \left(1 - \frac{k_z^2}{\omega^2} \right) |j_z|^2
%		= \left(1 - \cos^2\theta \right) |j_z|^2 \,.
        = \sin^2\theta \ |j_z|^2 \,.
\end{equation}

As we can see, the last line in Eq.~\eqref{Etot_beta_transf_2} is exactly the formula Eq.~\eqref{Etot_pointcharge} for the point charge,
\begin{equation}
	E_\text{tot}^\text{pt} = e^2 E_\text{tot}^\beta \,.
\label{sameE}
\end{equation}
Moreover, since this equality holds for \textit{any} rectilinear trajectory, we must have the correspondence not just between the energies but between the energy densities.
This correspondence can be read off from the derivation that we just presented, see Appendix~\ref{sec:recipes}.

\subsection{Application: specific trajectory}

Now let us see how this general correspondence works in particular case with the trajectory from Eq.~\eqref{traj_t}.
We employ the following convention for the Fourier transform:
\begin{equation}
	j_\mu(\omega, \bd{k}) = \int \diff^4 x j_\mu (t, \bd{x}) e^{-i k \cdot x} \,,
\label{fourier_def}
\end{equation}
where $x_\mu = (t, \bd{r})$, $k_\mu = (\omega, \bd{k})$ and $k \cdot x = k_\mu x^\mu =  \omega t - (\bd{k} \cdot \bd{r})$.

\paragraph{Fourier transforms}
Let us start from the Fourier transform of the three dimensional part of the current $\bd j$. The delta functions allows us to perform the spatial integration, and arrive at
\begin{equation}
	\bd j(\omega, \bd{k}) =  e\, \int\limits_{- \infty}^{\infty} \diff{t} \bd{v}(t) e^{-i (\omega t - \bd{k} \cdot \bd{r}(t)) } \,.
\end{equation}
On our trajectory Eq.~\eqref{traj_t} 
\begin{equation}
\begin{aligned}
	j_z (\omega, k_z) 
		&= e\, \int\limits_{- \infty}^{\infty} \diff{t} \dv{z}{t} e^{-i (\omega t - k_z z(t)) } \\
		&=  e\,  \int\limits_{0}^{\infty} \diff{z} e^{-i (\omega t(z) - k_z z) } \,.
\end{aligned}
\label{jz_fourier_transform_definition}
\end{equation}
Note the $\diff r $ integration in the second line, and the fact that for such rectilinear trajectory the current does not depend on $k_x$, $k_y$.
Substituting the trajectory from Eq.~\eqref{traj_t} we arrive at
\begin{equation}
	j_z (\omega, k_z) =  e\, \int\limits_{0}^{\infty} \diff{z} \left( \frac{\kappa z}{s} \right)^{-i \frac{\omega}{\kappa} } e^{-i z \left( \frac{\omega}{s}  - k_z \right) } \,.
\end{equation}
This integral is computed in Appendix~\ref{sec:int_calc}, the result is
%To compute the integral we need to transform it to the $\Gamma$-function form. 
%Suppose first that $\omega - k_z s > 0$.
%We make a change of variables (cf. footnote on p. \ref{ftnt:log})
%\begin{equation}
%	r = \rho \, \frac{e^{- i \frac{\pi}{2}} }{\omega / s - k_z}
%\end{equation}
%%
%and then rotate the contour of integration to the real axis. This yields
\begin{equation}
	j_z (\omega, k_z) = 
		 e \left( \frac{\kappa }{\omega  - k_z s} \right)^{-i \frac{\omega}{\kappa} }
		\frac{s e^{- i \frac{\pi}{2} - \frac{\pi \omega}{2 \kappa} } }{\omega  - k_z s}
		\Gamma \left( 1 - i \frac{\omega}{\kappa} \right) \,.
\label{j_omega_k}
\end{equation}
%
%This result can be also analytically continued to the region $\omega - k_z s < 0$.
%Using the identity $\Gamma(1-z) \Gamma(z)=\frac{\pi}{\sin \pi z}$ we can compute the absolute value
%\begin{equation}
%	| j_z (\omega, \bd{k}) |^2 = 
%		\frac{e^2 s^2 e^{- \frac{\pi \omega}{ \kappa}}}{(\omega  - k_z s)^2} \frac{\pi \omega / \kappa}{\sinh(\frac{\pi \omega}{\kappa}) } 
%\end{equation}
%%
%This can be simplified as
The absolute value squared is easily computed, see also Eq.~\eqref{fourier_abs_result},
\begin{equation}
	| j_z (\omega, k_z) |^2 = 
	\frac{2 \pi  s^2 e^2 }{ (\omega  - k_z s)^2} \frac{|\omega| / \kappa}{ e^{\frac{2\pi |\omega|}{\kappa}} - 1 }  \,.
\label{abs_j_omega_k}
\end{equation}
This formula is valid for $|\omega| \geqslant |k_z|$ (this is the case of interest, cf. the discussion above).
%
%What is the answer if $\omega - k_z s < 0$? 
%If we write $\omega - k_z s = |\omega - k_z s| e^{i \alpha}$ and trace back our calculation, we can see that for $\alpha \in [- 3 \pi / 2, \pi / 2]$ the same change of variables is still valid with the result
%\begin{equation}
%	| j_z (\omega, \bd{k}) |^2 = 
%	\frac{2 \pi  e^2 s^2 }{ (\omega  - k_z s)^2} \frac{e^{- 2 \alpha \omega} \,\omega / \kappa }{ e^{\frac{2\pi \omega}{\kappa}} - 1 } 
%\label{j_anyphase}
%\end{equation}
%%
%In particular, for $\omega - k_z s < 0$ we set $\alpha = - \pi$ and arrive at
%\begin{equation}
%	| j_z (\omega, \bd{k}) |^2 = 
%	\frac{2 \pi  e^2 s^2 }{ (\omega  - k_z s)^2} \frac{\omega / \kappa }{1 - e^{\frac{- 2\pi \omega}{\kappa}}  } 
%\end{equation}
%%
%Below we will be interested only in the case $|\omega| \geqslant |k_z|$. Since $1 \geqslant s$, this means in particular that the sign of $\omega - k_z s$ is the same as the sign of $\omega$, and we can write a more general formula:
%\begin{equation}
%	| j_z (\omega, \bd{k}) |^2 = 
%	\frac{2 \pi  e^2 s^2 }{ (\omega  - k_z s)^2} \frac{|\omega| / \kappa}{ e^{\frac{2\pi |\omega|}{\kappa}} - 1 } 
%	\label{abs_j_omega_k}
%\end{equation}

%\vspace{10pt}

The Fourier transform of the charge density can be computed in a similar fashion:
% We have:
%\begin{equation}
%	 \rho (\omega, \bd{k}) = e \int\limits_{- \infty}^{\infty} dt  e^{-i (\omega t - \bd{k} \cdot \bd{r}(t)) }
%	 	= e \int\limits_{0}^{\infty} dr \dv{t}{r} e^{-i (\omega t(r) - k_z r) }
%\label{rho_transform}
%\end{equation}
%%
%The reciprocal of the velocity is
%\begin{equation}
%	\dv{t}{r} = \frac{1}{\kappa r} + \frac{1}{s}
%\end{equation}
%%
%Substituting this back into the integral and assuming $\omega - k_z s > 0$, we can do the integration and compute the answer:
\begin{equation}
	\rho (\omega, k_z) = 
		 e \left( \frac{\kappa }{\omega  - k_z s} \right)^{-i \frac{\omega}{\kappa} }
		\frac{k_z s e^{- i \frac{\pi}{2} - \frac{\pi \omega}{2 \kappa} } }{\omega(\omega  - k_z s)}
		\Gamma \left( 1 - i \frac{\omega}{\kappa} \right) \,.
\label{rho_omega_k}
\end{equation}
The absolute value is
\begin{equation}
	| \rho (\omega, k_z) |^2 = 
		\frac{k_z^2}{\omega^2}
		\frac{2 \pi  s^2  e^2 }{ (\omega  - k_z s)^2} \frac{|\omega| / \kappa}{ e^{\frac{2\pi |\omega|}{\kappa}} - 1 } \,.
\label{rho_abs}
\end{equation}
Note that Eq.~\eqref{j_omega_k} and Eq.~\eqref{rho_omega_k} satisfy the charge conservation law Eq.~\eqref{cont_eq}, which is a good check.
Using our Fourier transforms from Eq.~\eqref{abs_j_omega_k} and Eq.~\eqref{rho_abs} we can write down the scalar Eq.~\eqref{j_scalar_omegapos}:
\begin{multline}
	\frac{|j_\mu (\omega, k_z)|^2 + |j_\mu (- \omega, k_z)|^2}{(2\pi)^2 e^2 } \\
		= \left( 1 - \frac{k_z^2}{\omega^2} \right)
		\frac{4 \pi  s^2 (\omega^2  + k_z^2 s^2) }{ (\omega^2  - k_z^2 s^2)^2} \frac{\omega / \kappa}{ e^{\frac{2\pi \omega}{\kappa}} - 1 } \,.
\label{j_pm_abs}
\end{multline}
Note that in this formula $\omega > 0$ is the physical frequency.

\paragraph{Connecting to the mirror}
Now we want to make a connection to the mirror moving in 1+1.
The beta Bogolubov coefficient for the corresponding mirror \cite{Good:2022eub} is
\begin{equation}
	|\beta_{pq}|^2 = \frac{2 s^2 p  q}{\pi  \kappa  (p+q) }\frac{ a^{-2} + b^{-2}}{ e^{2\pi (p+q)/\kappa}-1},
\label{betaT1}
\end{equation}
where $a= p(1+s) + q(1-s)$, and $b=p(1-s)+q(1+s)$.  

It can be easily checked that the change of variables Eq.~\eqref{pq_def} brings this beta Eq.~\eqref{betaT1} to the absolute squared current Eq.~\eqref{j_pm_abs}, thereby confirming our general formula Eq.~\eqref{beta_j_general}.
By doing explicit integrations we verified that the total radiated energies coincide, see Eq.~\eqref{sameE}.
We also checked that the algorithms presented in Appendix~\ref{sec:recipes} are consistent.

We checked this correspondence explicitly for several other trajectories as well.
Two important examples --- Schwarzschild and Carlitz-Willey --- are described in Appendix~\ref{sec:examples}.

\paragraph{An Energy-Zeta Function}
To relate the point charge and the mirror we used the light cone coordinates above, see Eq.~\eqref{pq_def}.
However there is another interesting set of coordinates which is also useful. 
Namely, consider new variables $m, \zeta$ defined as
\begin{equation}
\begin{aligned}
	2p &= \omega (1 + \cos\theta) = m e^\zeta,\quad \\
	2q &=\omega (1 - \cos\theta) = m e^{-\zeta},
\end{aligned}
\label{m_zeta_def}
\end{equation}
This transformation can be equivalently represented as
\begin{equation}
	\omega = p+q = m \cosh\zeta,\quad  
	k_z = p-q = m\sinh\zeta
\end{equation}
or as
\begin{equation}
	\zeta = -  \ln \tan \frac{\theta}{2} \,, \quad
	m = \omega \sin\theta \,.
\end{equation}
These variables are not new, see e.g. Eq.~(17) of \cite{Ritus:2003wu}, but we believe their usefullness might be underestimated in the literature.

\begin{figure}[htbp]
\centering
%\begin{subfigure}{0.5\textwidth}
  \centering
  \includegraphics[width=0.9\linewidth]{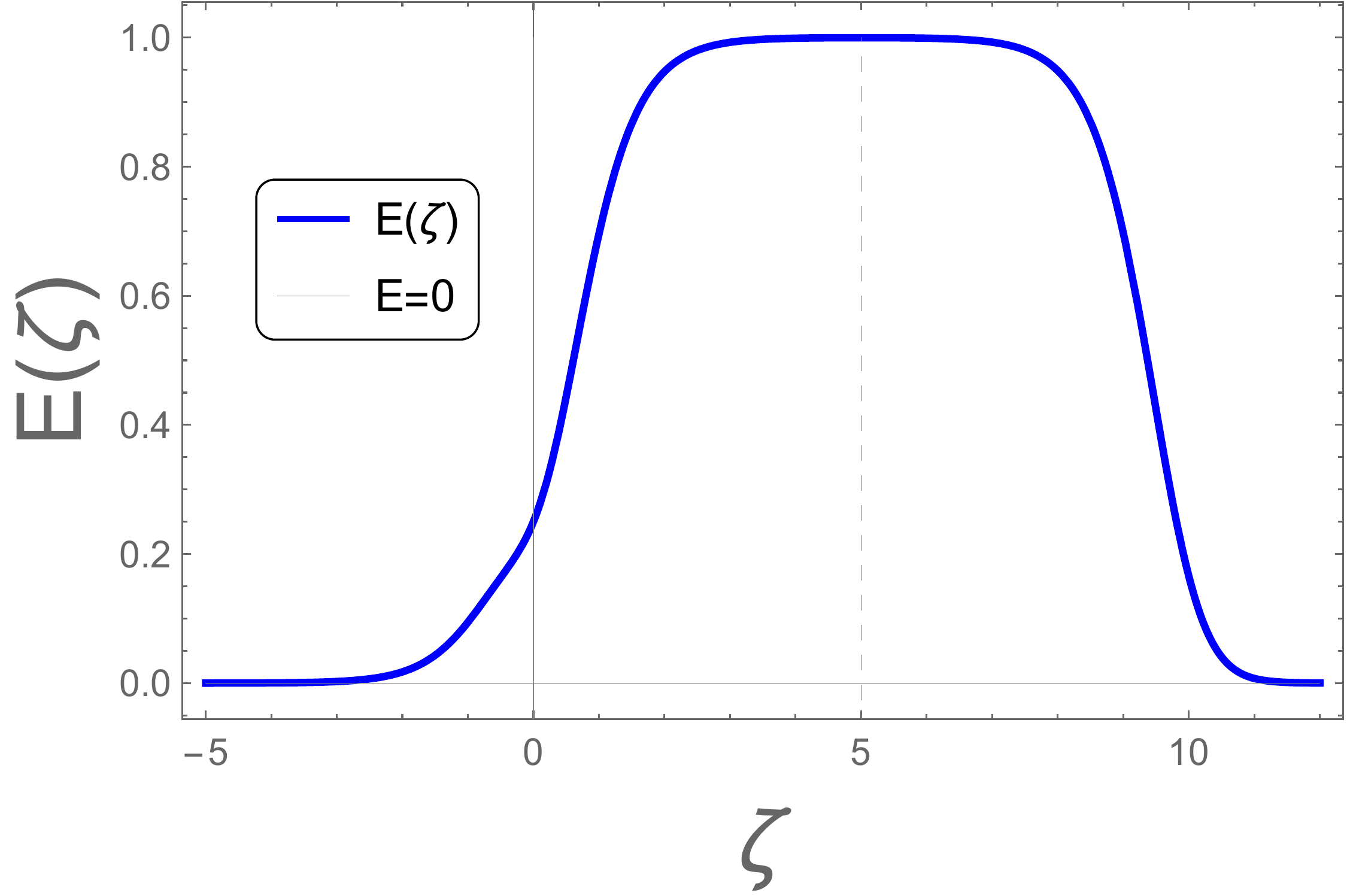}
 \caption{A plot of Eq.~(\ref{et}), the energy emitted by the mirror/electron as a function of $\zeta$, where $\kappa = 1$ and ultra-relativistic final speed $s=0.999,999,994$. The final speed is chosen so that the maximum $E_\textrm{max} = 0.999$ for the plateau which occurs at approximately $\zeta_0 = 5.00$; here $E(\zeta)$ is scaled by $24\pi$. The plateau is indicative of constant energy emission confirming a period of equilibrium radiation commensurate with Planck-distributed particles.}
\label{Fig1}
\end{figure}

An analytic \textquote{half-way} result is possible with this change of coordinates.   
The total energy emitted can be expressed as
\be E = \int p\; |\beta_{pq}|^2 \diff p \diff q, \ee
and using $ \diff p \diff q = (m/2)dm d\zeta = (1/4)dm^2 d\zeta$, we can integrate over $m$ from $0$ to $\infty$, using $\kappa = \mu_0 = c = 1$, 
\be E(\zeta) = \frac{s^2 e^{\zeta }\text{sech}^3\zeta }{48\pi} \frac{\left(\left(1+s^2\right) \cosh (2 \zeta )+1-s^2\right)}{ \left(\left(1-s^2\right) \cosh (2 \zeta )+1+s^2\right)^2},\label{et}\ee
and then integrating over $\zeta$  gives
\be E = \int_{-\infty}^{+\infty} E(\zeta) \diff \zeta,\ee
resulting in Eq.~(\ref{TE}). It is fortunate to have an analytic result for $E(\zeta)$, as opposed to the intractability of e.g. $E(p)$, $E(m)$, or $E(q)$.  
A plot demonstrating behavior indicative of thermal emission via a flatten energy plateau is illustrated in Figure \ref{Fig1}.

\section{Relevant Regimes}
\subsection{Thermal Limit}
In natural units, notice that our 1-D black body spectrum, Eq.~(\ref{I_result_1}),
\be	I(\omega) = \frac{e^2}{2\pi^2}\left(\frac{ \eta}{s}-1\right) \frac{2\pi \omega/\kappa}{e^{2\pi \omega/\kappa}-1},
 \ee
 is identically thermal in frequency $\omega$.  How can this be seen using the quantum analog moving mirror? or, namely using mode frequencies $p$ and $q$? %Using $\kappa = e= \mu_0 = c = 1$, the limit when $s\to 1$, of Eq.~(\ref{j2}), is
%\be \lim_{s\to 1}\frac{|j_\alpha(k)|^2}{(2\pi)^2} = \frac{1}{\pi \omega\left(e^{2 \pi  \omega }-1\right)}\frac{\omega ^2+k^2}{\omega ^2-k^2}.\ee
First we look at the limit of Eq.~(\ref{betaT1}) at high speeds, 
\be \lim_{s\to 1}|\beta_{pq}|^2 = \frac{p^2+q^2}{2 \pi \kappa  p q \left(e^{2 \pi  (p+q)/\kappa}-1\right) (p+q)}.\ee
Then, we follow Hawking \cite{Hawking:1974sw} and take what he called the `high frequency limit' of Eq.~(\ref{betaT1}), which is $q\ll p$. 
To lowest order in small $q = (\omega-k_z)/2$, the usual thermal result is obtained, e.g. Eq. 1 of Fulling \cite{Fulling_optics}
\be |\beta_{pq}|^2 = \frac{1}{2 \pi \kappa q}\frac{1}{  e^{2 \pi  p/\kappa}-1 }.\ee
This formula is well-known as the thermal Planck distribution associated with eternal black body emission, e.g. Carlitz-Willey \cite{carlitz1987reflections} or the DeSitter mirror \cite{Good:2020byh}.  It also appears at late-times for the aforementioned Schwarzschild \cite{Good:2016oey}, Reissner–Nordstr\"om \cite{good2020particle}, and Kerr \cite{Good:2020fjz} black hole mirror analogies.  Thus our result confirms thermal radiation in the ultra-relativistic limit, $s\to 1$ and the high frequency regime, $q\ll p$. 

\paragraph*{Interpretation}
We can offer the following interpretation in terms of the 3+1 electron.
As can be seen from Eq.~\eqref{dIdOmega}, the angular distribution is peaked at $\theta = \cos^{-1}(s) $, and in the ultrarelativistic limit $s \to 1$ most of the energy is radiated almost forward at small angles $\theta \sim \sqrt{2(1-s)}$.

To obtain the radiation spectrum $I(\omega)$ one must integrate the angular distribution over the solid angle. 
However at high final speeds the integral is saturated at small angles $\theta$, so that
\begin{equation}
	\frac{\diff I(\omega)}{\diff \Omega} \Big|_{\theta \sim 0} \sim I(\omega)
\label{I_ultrarel}
\end{equation}
up to a coefficient. 

Now let us look back at the mirror.
In terms of $\omega$ and $\theta$ (see Eq.~\eqref{m_zeta_def}) the condition $q\ll p$ becomes simply $\theta \ll 1$.
Therefore the ultrarelativistic mirror corresponds to a forward radiation of the ultrarelativistic electron.
Since $|\beta_{pq}|^2$ is directly related to $\diff I / \diff \Omega$ (see Appendix~\ref{sec:recipe_I2beta}), Eq.~\eqref{I_ultrarel} tells us that $|\beta_{pq}|^2$ is proportional to $I(\omega)$ in this limit, i.e. it gives the thermal Planck distribution.

\subsection{Infrared Limit}
Our electron emits soft particles which have long wavelengths that lack the capability to probe the internal structure of their source \cite{Strominger:1994tn}.  Is the spectrum, $I(\omega)$, Eq.~(\ref{Iw}), consistent with the deep infrared?  

In the deep infrared limit of $\omega \to 0$, the spectrum, $I(\omega)$, Eq.~(\ref{I_result_1}), in SI units, becomes
\be I_{\textrm{infra}} = \frac{\mu_0 c e^2}{2\pi^2}\left(\frac{\tanh^{-1} s/c}{s/c}-1\right),\label{Iinfra}\ee
independent of frequency $\omega$. This agrees with the well-known frequency-independent deep IR result and characterizes the step-function trajectory, see e.g. Prob. 23.16 of Zangwell \cite{Zangwill:1507229}, Nikishov-Ritus Sec.~2 \cite{Nikishov:1995qs}, or Jackson Sec.~15.1A \cite{Jackson:490457}.
\subsection{Global Limit}
The total energy of the spectrum $I(\omega)$ is expected to be finite.  In the global limit, considering all the colors emitted, the total energy can be found by integrating $I(\omega)$ of Eq.~(\ref{I_result_1}) over $\omega$,
\be E = \int_0^\infty I(\omega) \diff \omega,\ee
which gives, in SI units,
%\be E = \frac{1}{24\pi}\left(\frac{\eta}{s}-1\right).\label{TE}\ee
%\be E = \frac{\mu_0 e^2\kappa}{24\pi}\left(\frac{c\eta}{s}-1\right).\label{TE}\ee
\begin{equation}
	E = \frac{\mu_0 e^2\kappa}{24\pi}\left(\frac{ \tanh^{-1}(s/c)}{s/c}-1\right).
\label{TE}
\end{equation}
This is Eq.~\eqref{TE0} and it agrees with the integration of the relativistic Larmor power over time \cite{Good:2022xin} see also \cite{Good:2022eub}.  This total energy has been experimentally confirmed; e.g. \cite{Ballagh:1983zr} and has long been associated with beta decay \cite{PhysRev.76.365}.  It is also the energy from the instantaneous collision formalism \cite{Cardoso:2002pa,Cardoso:2003cn}.

\section{Conclusions}

A classical computation reveals a moving point charge can emit identically thermal radiation. The process reveals the temperature is proportional to the acceleration.  The relevance to the Davies-Fulling-Unruh effect in quantum field theory reveals a correspondence between the electron and moving mirror. 

Compelling experimental confirmation is possible.  In particular; three prominent future avenues stand out: 
\begin{itemize}
\item Extreme accelerations experienced by an electron during the process of radiative free neutron beta decay \cite{nico}, in the RDKII collaboration experiment \cite{RDKII:2016lpd} provides an ideal system to examine the trajectory of this particular electron-mirror, see the recent analysis \cite{Lynch:2022rqx}.
    \item More generally, the Analog Black Hole Evaporation via Lasers (AnaBHEL) experiment \cite{AnaBHEL:2022sri} demonstrates that the relativistic trajectory \cite{Chen:2015bcg} of the  moving mirror is an important probe for the spectral physics of quantum vacuum radiation, via an accelerated plasma \cite{Chen:2020sir}.  
\item Moving mirror inspired metrics, e.g. \cite{Good:2020fsw} from \cite{Good:2019tnf}, enable experimental investigation and can provide high-resolution observational data using optical analogues via a continuous-wave a light field pulse filling the fibre \cite{Moreno-Ruiz:2021qrf}. %They are implemented in the laboratory by sending a light pulse of the desired shape through an optical fibre. Continuous-wave light fields can be used to fill the fibre and approximate event horizons \cite{Moreno-Ruiz:2021qrf}.
\end{itemize}
On the theory side, one potentially interesting future study would be the application of the above recipe to non-thermal asymptotic resting motions with finite particle count; e.g. \cite{Walker_1982}.

\begin{acknowledgments}
The authors thank V. Ritus, who at the age of 95, generously shared (solicited by us) several related papers. We are grateful for his contributions to the field and his willingness to provide his knowledge and expertise.  Funding comes in part from the FY2021-SGP-1-STMM Faculty Development Competitive Research Grant No. 021220FD3951 at Nazarbayev University. %Appreciation is given to the organizers, speakers, and participants of the QFTCS Workshop: May 23-27, 2022, at which preliminary results were first presented and helpful feedback are included therein.       
%\newpage
\end{acknowledgments}

\appendix

%\section{A Tale of Two Thermometers}

%Encountering the spectrum, Eq.~(\ref{Iw}),
%\be I(\omega) = \frac{\mu_0 c e^2}{2\pi^2}\left(\frac{c \eta}{s}-1\right) \frac{2\pi c \omega/\kappa}{e^{2\pi c \omega/\kappa}-1},\ee

\section{Technicalities of the electron-mirror correspondence}
\label{sec:recipes}

In this Appendix we present two algorithms for passing between a 1+1 mirror and a 3+1 point charge in a rectilinear motion.
These instruction manuals can be read off the derivation presented in Sec.~\ref{sec:to_the_mirror}.

%\subsection{Recipe: from $|\beta_{pq}|^2$ to $I(\omega)$}
%
%\begin{enumerate}
%	\item From the mirror take $|\beta_{pq}|^2$ and substitute $p = \frac{1}{2}(\omega + k_z)$, $q = \frac{1}{2}(\omega-k_z)$.
%	\item Integrate $\int\limits_{-\omega}^{+\omega} \diff k_z$.
%	\item Multiply by $\omega / 4$. The result is $I(\omega)$ --- radiated energy density from the accelerated electron.
%\end{enumerate}

\paragraph{Recipe: from $|\beta_{pq}|^2$ to $I(\omega)$}
%\subsection*{Recipe: from $|\beta_{pq}|^2$ to $I(\omega)$}
\label{sec:recipe_beta2I}

\begin{enumerate}
	\item On the mirror side take $|\beta_{pq}|^2$ and substitute $p = \frac{\omega}{2}(1 + \cos\theta)$, $q = \frac{\omega}{2}(1 - \cos\theta)$.
	\item Integrate $\int\limits_{-1}^{+1} \diff (\cos\theta) $.
	\item Multiply by $e^2 \omega^2 / 4$. The result is $I(\omega)$ --- radiated energy density, the `spectrum', from the accelerated electron.
\end{enumerate}

\paragraph{Recipe: from $\diff I / \diff \Omega$ to $|\beta_{pq}|^2$ }
%\subsection*{Recipe: from $\diff I / \diff \Omega$ to $|\beta_{pq}|^2$ }
\label{sec:recipe_I2beta}

Use the following formula:

\be |\beta_{pq}|^2 = \frac{4\pi}{e^2 \omega^2}\left[\frac{\diff{I}}{\diff{\Omega}}(\omega,\cos\theta) + \frac{\diff{I}}{\diff{\Omega}}(\omega,-\cos\theta)\right],\ee

where $p+q=\omega$ and $p-q = \omega \cos\theta$.

%\subsection{Recipe: from $\dv{I(\omega)}{\Omega}$ for a rectilinear motion to $|\beta_{pq}|^2$   }
%\label{sec:recipe_I2beta}

%\begin{enumerate}
%	\item Express $\dv{I(\omega)}{\Omega}$ as a function of $\omega$ and $\cos\theta$
%	\item Substitute $\cos\theta = k/\omega$
%	\item Multiply by $4\pi/\omega$
%	\item Take a sum with $+k$ and $-k$ (i.e. $(p,q) + (q,p)$)
%	\item Substitute $\omega= p+q$, $k=p-q$
%\end{enumerate}

%\begin{enumerate}
%	\item On the point charge side take $\dv{I(\omega)}{\Omega}$ and express it as a function of $\omega$ and $\cos\theta$
%	\item Substitute $\omega = p+q$, $\cos\theta = \frac{p-q}{p+q}$. The result is some function, let's denote it $f(p,q)$.
%	\item Compute the beta Bogolubov as $\frac{4\pi}{(p+q)^2} (f(p,q) + f(q,p)) \equiv |\beta_{pq}|^2$
%	\item Formula $\frac{4\pi}{(p+q)^2} (f(p,q) + f(q,p)) \equiv |\beta_{pq}|^2$ gives the beta Bogolubov.
%\end{enumerate}
%\mike{maybe this should be done for uniform acceleration? standard rectilinear bremsstrahlung \cite{Griffiths:1492149}; what mirror betas result? it should agree with the usual uniformly accelerated mirror betas ; e.g. \cite{good2020extreme}}

\section{Further examples of the electron-mirror correspondence}
\label{sec:examples}

In this Appendix we present two further examples of this correspondence, the Schwarzschild and Carlitz-Willey trajectories.
These examples are qualitatively different from the trajectory in Eq.~\eqref{traj_t} in that the electron now approaches the speed of light. (On the mirror side the 1+1 spacetime develops a horizon.)

The electron radiation flux becomes singular at certain angles. 
The spectral distribution $\diff I(\omega) / \diff \Omega$ exists, but the integrated quantity $I(\omega)$ is in fact divergent.
Still, we can investigate the angle-dependent spectral distribution.

We find that in the receding-redshift limit $\diff I(\omega) / \diff \Omega$ is thermal with temperature given by the same Eq.~\eqref{Tnatural}.
This means that the observer behind the electron will see a redshifted thermal radiation.

%\subsection*{Schwarzschild trajectory}
\paragraph{Schwarzschild trajectory}
The so-called Schwarzschild mirror investigated in \cite{Good:2016oey} is a special case of a moving mirror.
The spacetime with such a mirror develops a horizon at late times. This system is directly related to the Schwarzschild black hole, see e.g. \cite{Good:2018zmx,Cong:2018vqx,wilczek1993quantum}.

The mirror's trajectory for this case can be cast in the form (late-time motion defined to be in the $+\bd{\hat{z}}$ direction)
\begin{equation}
	t(z) =  z - 4 M e^{- z / 2M} \,.
\end{equation}
The Fourier transform of the current Eq.~\eqref{jz_fourier_transform_definition} can be calculated by passing to the variable $h = e^{-z/2M}$ and then applying the result of Appendix~\ref{sec:int_calc}.
The angular power distribution is calculated using Eq.~\eqref{I_Omega}:
\begin{equation}
	\dv{I(\omega)}{\Omega} = 
%		\frac{ M  \omega (1 - \cos\theta) }{ 4 \pi^2 } \frac{1}{e^{4 \pi M \omega (1 + \cos\theta)} - 1}
		\frac{  \omega (1 + \cos\theta) }{ 16 \pi^2 \kappa } \frac{1}{e^{ \pi  \omega (1 - \cos\theta) / \kappa} - 1} \,,
\label{dIdOmega_schw}
\end{equation}
where $\kappa = 1/4M$ (the analog of \textquote{surface gravity}). 
In the receding-redshift limit $\theta \sim \pi$ the distribution Eq.~\eqref{dIdOmega_schw} reduces to a 1+1 Planck distribution.
The temperature in terms of the parameter $\kappa$ is given by exactly the same expression as Eq.~\eqref{Tnatural}.

The recipe from Appendix~\ref{sec:recipes} indeed gives the beta Bogolubov for the Schwarzschild mirror (both sides of the mirror)
\begin{equation}
	|\beta_{pq}|^2 = \frac{1}{ 2 \pi \kappa (p+q)^2 } \left( \frac{q}{e^{ 2 \pi  p / \kappa} - 1} + \frac{p}{e^{ 2 \pi  q / \kappa} - 1} \right) \,.
\end{equation}
For the background on this mirror see \cite{Good_2017Reflections1}.
%This really does look like left + right side of the mirror, cf. Eq.~(28) of \cite{Good:2018zmx} or Eq.~(22) of \cite{Good_2017Reflections1}.

%\subsection*{Carlitz-Willey trajectory}
\paragraph{Carlitz-Willey trajectory}

Another case of interest is the trajectory first studied by Carlitz-Willey \cite{carlitz1987reflections,CW2lifetime}, 
\begin{equation}
	z(t) = t + \frac{1}{\kappa} W\left(e^{-2 \kappa t}\right) \,,
\end{equation}
where $W$ is the Lambert product logarithm, with the above transcendental form introduced in \cite{Good:2012cp} and studied, e.g. in \cite{good2013time,Akal:2021foz,Good:2022wpw,Akal:2022qei}. Late-time motion is in the $+\bd{\hat{z}}$ direction. 

The Fourier integral in Eq.~\eqref{jz_fourier_transform_definition} is computed with the help of the variable $h = W( e^{-2 \kappa t} )$ and Appendix~\ref{sec:int_calc}.
The result for the angular power distribution from Eq.~\eqref{I_Omega} is
\begin{equation}
	\dv{I(\omega)}{\Omega} = 
		\frac{\omega}{ 4 \pi^2 \kappa (1 + \cos\theta) } \frac{1}{e^{\pi \omega (1 - \cos\theta) / \kappa} - 1} \,.
\label{dIdOmega_CW}
\end{equation}
In the receding-redshift limit $\theta \sim \pi$ the distribution Eq.~\eqref{dIdOmega_schw} reduces to 1+1 Planck. The temperature in terms of the parameter $\kappa$ is given by exactly the same expression as Eq.~\eqref{Tnatural}.

The recipe from Appendix~\ref{sec:recipes} indeed gives the beta Bogolubov for the Carlitz-Willey mirror (both sides)
\begin{equation}
	|\beta_{pq}|^2 = \frac{1}{ 2 \pi \kappa } \left(  \frac{q^{-1}}{e^{\frac{2 \pi p}{ \kappa}} - 1} +  \frac{p^{-1}}{e^{\frac{2 \pi q}{ \kappa}} - 1} \right) \,.
\end{equation}

\section{Calculation of Fourier integrals}
\label{sec:int_calc}

In this section we will derive a general formula for Fourier transforms that are used in this paper,
\begin{equation}
	A(\alpha, \beta, \gamma) = \int\limits_0^\infty  z^{i \alpha + \beta} e^{i \gamma z} \diff z
\end{equation}
%
%with $\alpha, \gamma > 0$ and $\beta \in \mathbb{R}$.  
with $\alpha, \beta, \gamma \in \mathbb{R}$ and $\alpha \cdot \gamma > 0$.
For the complex multivalued function $z^{i \alpha + \beta}$ we choose the standard Riemann sheet with $1^{i \alpha + \beta}=1$ and the cut going from $0$ to $-\infty$.
To calculate this integral we make a change of variables from $z$ to $\rho$
\begin{equation}
	z = \rho \,  \frac{e^{i \frac{\pi}{2} \operatorname{sign}(\gamma)  }}{|\gamma|}
\end{equation}
and rotate the contour of $\rho$-integration to the real axis, which yields
\begin{equation}
	A(\alpha, \beta, \gamma) = \frac{ e^{- |\alpha| \frac{\pi}{2} + i \frac{\pi}{2} (\beta + 1) \operatorname{sign}(\gamma) } }
			{|\gamma|^{i \alpha + (\beta + 1)} }
			\int\limits_0^\infty \diff\rho \rho^{i\alpha + \beta} e^{- \rho} \,.
\end{equation}
This is an integral representation of the $\Gamma$-function, so that
\begin{equation}
	A(\alpha, \beta, \gamma) = \frac{ e^{- |\alpha| \frac{\pi}{2} + i \frac{\pi}{2} (\beta + 1) \operatorname{sign}(\gamma) } }
				{|\gamma|^{i \alpha + (\beta + 1)} }
			\Gamma(i\alpha + (\beta+1) ) \,.
\end{equation}
Using the identity
\begin{equation}
	\Gamma(1-z) \Gamma(z) = \frac{\pi}{\sin \pi z}
\end{equation}
we can compute 
\begin{equation}
	|\Gamma(i\alpha)|^2 = \frac{\pi}{\alpha \sinh\pi\alpha } \,, \quad
	|\Gamma(i\alpha + 1 )|^2 = \frac{\pi\alpha}{ \sinh\pi\alpha } \,.
\end{equation}
This gives two important cases:
%\begin{equation}
%\begin{aligned}
%	|A(\alpha, -1, \gamma)|^2 &= \frac{2\pi  }{\alpha} \frac{1}{e^{2\alpha \pi} - 1} \\
%	|A(\alpha, 0, \gamma)|^2 &= \frac{2\pi\alpha  }{\gamma^2} \frac{1}{e^{2\alpha \pi} - 1}
%\end{aligned}
%\end{equation}
%
\begin{equation}
\begin{aligned}
	\left| \int\limits_0^\infty  z^{i \alpha - 1} e^{i \gamma z} \diff z \right|^2 &= \frac{2\pi  }{|\alpha|} \frac{1}{e^{2|\alpha| \pi} - 1} \,, \\
	\left| \int\limits_0^\infty  z^{i \alpha } e^{i \gamma z} \diff z \right|^2 &= \frac{2\pi|\alpha|  }{\gamma^2} \frac{1}{e^{2|\alpha| \pi} - 1} \,.
\end{aligned}
\label{fourier_abs_result}
\end{equation}
Note that the first expression does not actually depend on $\gamma$.

\bibliography{main} 
\end{document}